\documentclass[journal]{IEEEtran}
\usepackage{cite}
\usepackage{amsmath,amssymb,amsfonts}
\usepackage{algorithm}
\usepackage{listings}
\usepackage{graphicx}
\usepackage{textcomp}
\usepackage{algpseudocode} 
\usepackage{float} 
\usepackage{xcolor}
\usepackage{enumitem}
\usepackage{tabularx}
\usepackage[utf8]{inputenc}
\usepackage{array}
\usepackage{comment}
\usepackage{caption}
\usepackage{subfig}
\usepackage{textcomp}
\usepackage{mathtools}
\usepackage{hyperref}

\graphicspath{ {./figures/} }

\begin{document}

\title{Multiplier-free In-Memory Vector-Matrix Multiplication Using Distributed Arithmetic}

% \author{\IEEEauthorblockN{Felix Zeller}
% \IEEEauthorblockA{\textit{Friedrich-Alexander-Universit\"at \\
% Erlangen-N\"urnberg (FAU)\\
% felix.zeller@fau.de}}
% \and
% \IEEEauthorblockN{John Reuben}
% \IEEEauthorblockA{\textit{Friedrich-Alexander-Universit\"at \\
% Erlangen-N\"urnberg (FAU)\\
% johnreuben.prabahar@fau.de}}
% \and
% \IEEEauthorblockN{Dietmar Fey}
% \IEEEauthorblockA{\textit{Friedrich-Alexander-Universit\"at  \\
% Erlangen-N\"urnberg (FAU)\\
% dietmar.fey@fau.de}}
% }
\author{Felix Zeller, John Reuben and Dietmar Fey 
 \thanks{\textbf{Both Felix Zeller and John Reuben contributed equally to this work}}      
\thanks{Authors are with Chair of Computer Architecture, Friedrich-Alexander-Universit\"at  Erlangen-N\"urnberg (FAU), 91058 Erlangen, Germany. (email:johnreubenp@gmail.com )}% <-this % stops a space
%% <-this % stops a space
}

\maketitle

%\IEEEoverridecommandlockouts
%\IEEEpubid{\makebox\[\columnwidth\]{$^{*}$\textbf{Both Felix Zeller and John Reuben contributed equally to this work}}}
%\hspace{\columnsep}\makebox[\columnwidth]{ }
%\footnote{Both Felix Zeller and John Reuben contributed equally to this work}

%\IEEEpubidadjcol

\begin{abstract}
Vector-Matrix Multiplication (VMM) is the fundamental and frequently required computation in inference of Neural Networks (NN). Due to the large data movement required during inference, VMM
can benefit greatly from in-memory computing. However, ADC/DACs required for in-memory VMM consume significant power and area. `Distributed Arithmetic (DA)', a technique in computer architecture prevalent in 1980s was used to achieve inner product or dot product of two vectors without using a hard-wired multiplier when one of the vectors is a constant. In this work, we extend the DA technique to multiply an input vector with a constant matrix. By storing the sum of the weights in memory, DA achieves VMM using shift-and-add circuits in the periphery of ReRAM memory. We verify functional and also estimate non-functional properties (latency, energy, area) by performing transistor-level simulations. Using energy-efficient sensing and fine grained pipelining, our approach achieves 4.5 $\times$ less latency and 12 $\times$ less energy than VMM performed in memory conventionally by bit slicing. Furthermore, DA completely eliminated the need for power-hungry ADCs which are the main source of area and energy consumption in the current VMM implementations in memory.
\end{abstract}


%%%%%%%%%%%%%%%%%%%%%%%%%%%%%%%%%%%%%%%%%%

\section{Introduction}
\par  Due to  a phenomenon called `von Neumann bottleneck' (also called `memory wall'), conventional computer architecture is being re-engineered. The memory wall problem has two facets: the mismatch in the performance (speed) of processor and memory and the energy for data transfer during memory access. The energy to access (transfer) data is growing exponentially along the memory hierarchy (from cache to off--chip DRAM to tertiary Non Volatile Memory (NVM)). Consequently, `data movement energy' dominates the `computation energy'  in traditional systems with separate memory and processing units, $i.e.$ the computation in itself consumes only a small fraction of the energy \cite{powerwallref,patmos}. This becomes more severe with the advent of AI. Neural Networks typically require processing on large amounts of data and von Neumann architecture will be inefficient for such AI tasks since large amount of data have to be moved frequently between memory and processor. Inference of Convolutional Neural Network (CNN) is one such AI task frequently used in computer vision and autonomous driving. Even small CNNs perform millions of Multiply and Accumulate (MAC) operations per layer and conventional von Neumann architecture will be inefficient for such processing if the input image and weights have to be shuttled between processor and memory.
\par To address these challenges, academia and industry have been exploring alternate paradigms in the recent years. In-Memory Computing(IMC) is one such paradigm in which the memory and processing are deliberately brought together to make computing energy efficient. Computing is performed in the memory array or in the peripheral circuitry or in computing units placed near the memory array (also called near-memory computing). The demarcation between In-Memory Computing and Near-Memory Computing is blurred since the field is evolving fast and many architectures are being proposed with computation shared between the memory array and the circuitry around it, with varying degree of sharing \cite{patmos}. Vector Matrix Multiplication (VMM) is the fundamental and frequently required operation in neural networks, image processing, combinatorial optimization, solving linear equations, sparse coding, associative memories, reservoir computing and many signal processing applications \cite{inmemory2,mvmreview}. According to \cite{aiaccelerator,allanalog} VMM operations constitute 90 \% of the workloads for inference of Deep Neural Networks (DNNs). Hence, architectures which can execute VMMs efficiently are of paramount importance.

\begin{figure}[h]
\centering
\includegraphics[scale=0.45]{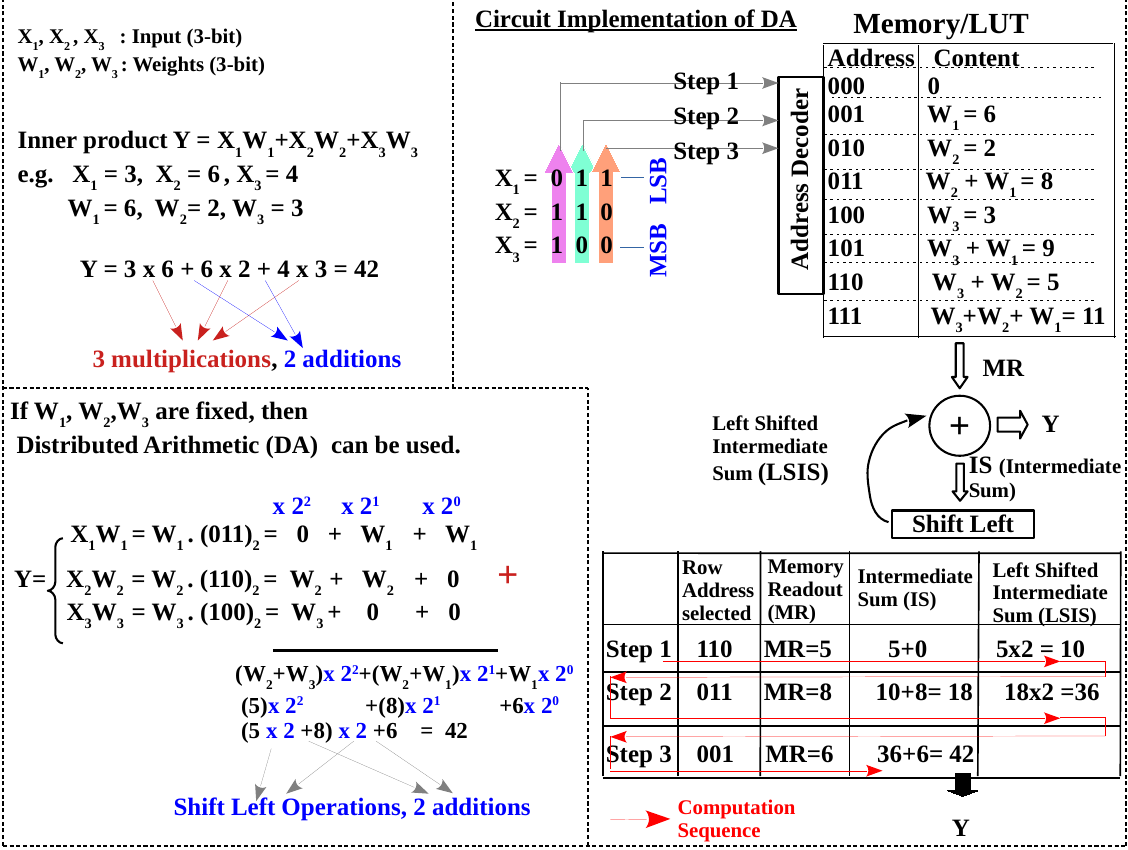}
\caption{\small{Simplified illustration of Distributed Arithmetic}}
\label{daillustration}       % Give a unique label
\end{figure}

\par Existing IMC architectures mostly implement VMM in an analog manner-- the elements of the matrix are programmed as conductance of the memory array and the elements of the vector are applied as voltages at the rows through a DAC. There exists two problems with existing IMC architectures. First, ADC consume huge energy and  occupy large area of the memory array. It is estimated that ADCs consume $\approx$ 60-85\% of the total energy and 70-90\% of the total area of a VMM core based on NVM arrays. \cite{mvmreview,pimbookchapter,imcbitslicing,adcless}.  Secondly, programming the memory cell to multiple states is a challenge in most non-volatile (Flash, ReRAM, FeFET) memory technology\footnote{In both ReRAM and flash memory, programming to more than two states is achieved by write-verify algorithm where the memory cell is programmed with a series of pulses with progressively increasing amplitude to overcome variability. This is costly in terms of time and energy. }. If a 5-bit weight has to be stored in memory, we need a memory cell which can store 32 different states. The second problem is overcome by a technique called bit slicing \cite{issac,imcbitslicing}. In this approach, the 5-bit weight is stored as a binary number in 5 different columns of the array and VMM is achieved by shifting the digital results of a column (to implement $\times$ $2^{x}$) and adding them. This increases the latency when compared to pure analog VMM, but eliminates the need for the memory cell to be programmed to multiple states (which is difficult to achieve and also complicates the peripheral circuitry). However, even bit slicing needs ADC and the ADC resolution increases with increase in the number of rows of the matrix. In this work, we overcome both these problems (requirement of ADC and multi-bit programming of NVM cell) using an old technique called Distributed Arithmetic.

% fig 1

Distributed Arithmetic (DA) is an efficient procedure for computing inner product between a fixed vector and a variable vector. DA was proposed as early as 1974 and was an extensive area of research in the 1980-1990s \cite{whiteDAsurvey}.  The inner product of two vectors is the sum of the products of the corresponding components of these vectors.
\begin{equation} \label{eq1}
    Y = X^TW =  \sum_{i=1}^{N}X_iW_i
\end{equation}
where $X^T$ = [$X_1,X_2,...,X_{N}$] is the variable vector and
 $W^T$ = [$W_1,W_2,...,W_{N}$] is a fixed vector.
This is the typical computing requirement in neural network `inference' $e.g.$ [$W_1,W_2,...,W_{N}$] could be weights in neural networks and they remain constant throughout the `inference' (Once the Neural Network is trained, the weights do not change and they can be considered as constants). Conventionally, this inner product is computed using Multiply and Accumulate (MAC) unit which multiplies weight $W$ with the input vector $X$ and adds them to the partial product to get the final sum \cite{meher2017}.  DA is an alternative technique to compute the same inner product $ Y = A^T\cdot X$ without using multipliers. As depicted in Fig. \ref{daillustration}, the sum of weights are pre-computed and stored in the memory as a look-up Table. The input vector $X$ is represented as bits and as depicted in Fig.\ref{daillustration}, one bit of $X_{1},X_{2},X_{3}$ is fed to the memory array in a bit serial manner. This forms the address and the corresponding data in the memory is read out in steps 1--3. In each step, the read out $MR$ is left-shifted (to achieve $\times$2) and added to cumulatively compute $Y$. In this manner, DA achieves a multiplier-less architecture when one of the vectors is a constant. The biggest advantage of DA is the elimination of hard-wired multiplier unit since multiplication operation is energy and latency consuming in both CMOS-based ASIC implementation and in-memory implementation  \cite{wallacetree}.
\par In this work, we extend this DA technique which was originally proposed for Vector-Vector Multiplication (Inner Product) to Vector-Matrix Multiplication (VMM). Our significant contributions can be summarized as follows.\\
a) We propose an innovative method which exploits DA to perform VMM in memory without power hungry and area-consuming ADC/DACs.\\
b) We verify the proposed innovation by transistor-level simulation of memory array and the peripheral circuits.\\
c) Using fine-grain pipelining and low energy sensing technique, we achieve VMM in memory which is both fast and energy-efficient compared to performing the same VMM by bit-slicing.\\ 
The rest of the paper is organized as follows.  Section \ref{vmmbyda} presents our architecture to perform VMM in memory using DA. Section \ref{circuitimplementationsection}  presents design of sensing circuit and the computation units (adders, shifter) around the memory array to perform VMM. After verifying by transistor-level simulation, we also present non-functional properties (latency, energy) of VMM approach. In Section \ref{comparisonsection}, we compare our In-Memory VMM method with conventional bit-slicing approach and highlight the benefits of our approach. Section \ref{conclusionsection} concludes our work.

% fig 2
\begin{figure}[h]
\centering
\includegraphics[scale=0.6]{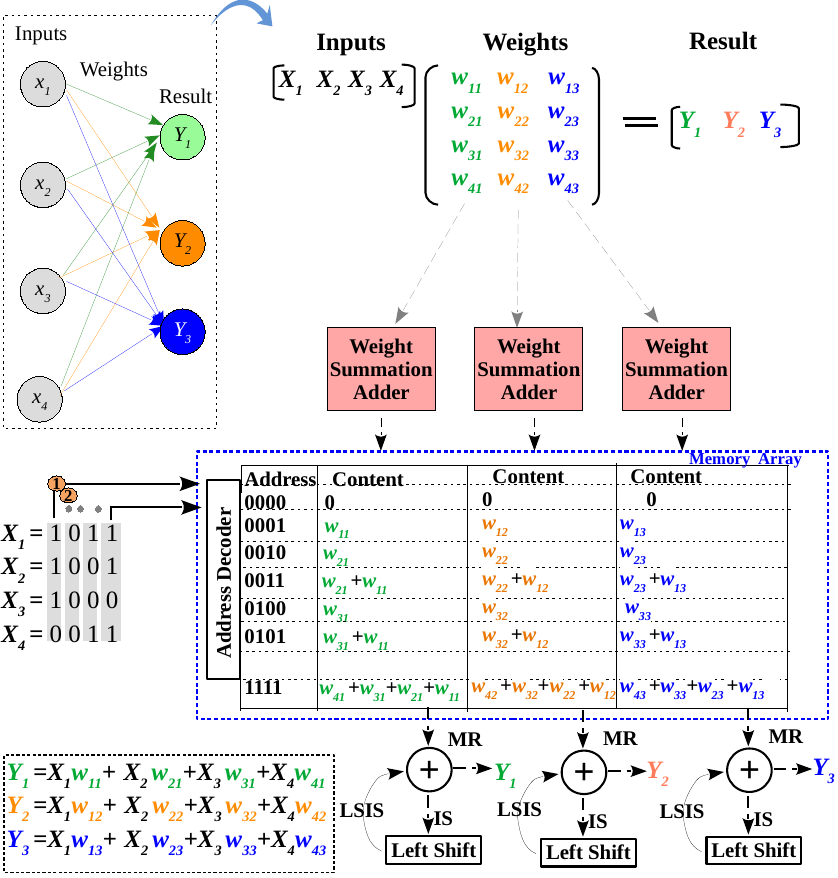}
\caption{\small{The weights of the full connected NN forms the matrix $W$ and the inputs form vector $X$. The sum of the weights is written into the processing memory and $X$ is applied in a bit-serial manner.}}
\label{mvmusingda}       % Give a unique label
\end{figure}

\section{VMM by Distributed Arithmetic (DA)}
\label{vmmbyda}
\subsection{Principle of VMM using DA}
VMM can be viewed as many inner-product computations  and they can be implemented simultaneously in the columns of a memory array. Fig.\ref{mvmusingda} depicts the multiplication of vector $X$ with a 4$\times$3 weight matrix $W$. As depicted in Fig.\ref{mvmusingda}, by having adders in the periphery of the memory array, we can essentially perform VMM in memory in a bit serial manner. The weights are first summed to calculate all possible sum of the weights by the `weight summation' adders and then written to the memory array. Once the sum of the weights are written, the input vector is sliced and in each cycle, 1-bit of $X_{1},X_{2},X_{3},X_{4}$ is applied to the array's address decoder. The decoder decodes the binary address and outputs the value at the corresponding address (`MR' denoting Memory Readout). This is fed to an adder which accumulates and adds it with the left-shifted MR value to compute $Y_{1},Y_{2},Y_{3}$. In this manner, VMM is achieved without a multiplier (thanks to DA). One must observe that addition, in principle can be performed in the memory array itself and several in-memory adders have been proposed in the recent past (\cite{magicbased,iptcas1,tvlsi}). However, the fastest method to add two $n$-bit numbers `in-memory' requires 4+6$\cdot$ $log _{2}n$ cycles (\cite{tvlsi}) while using CMOS adders `near the memory', one can perform such an addition in 1 cycle. Hence, in this work, we use CMOS adders in the periphery of the memory array since we want to minimize latency and ideally achieve latency comparable to application-specific hardware for VMM.

\begin{figure}[h]
\centering
\includegraphics[scale=0.6]{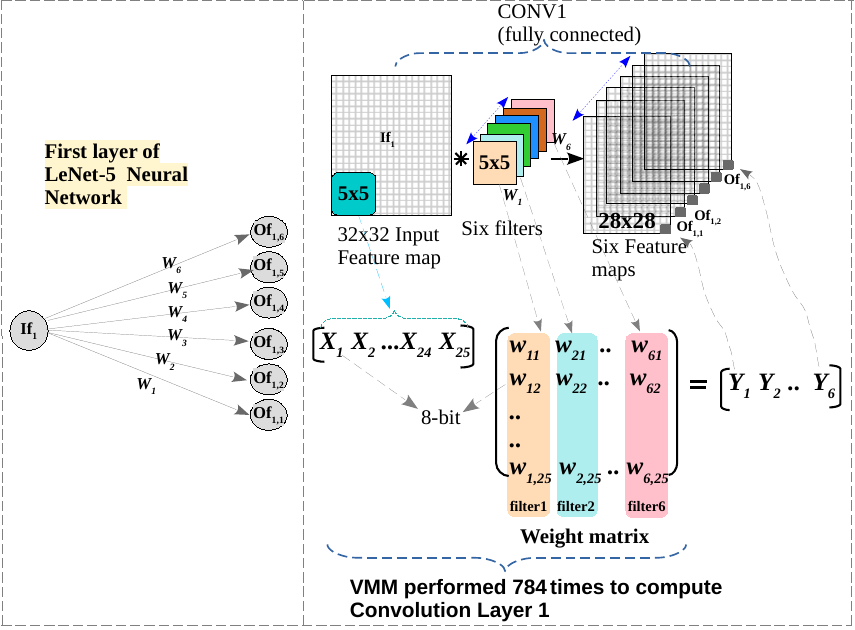}
\caption{\small{Illustration of mapping of the frist convolutional layers of LeNet-5 to Vector and Matrices. $If_{1}$ and $Of_{1,i}$ represent the input feature maps and output feature maps of convolutional layer 1. Each stride of the convolution becomes a VMM.}}
\label{mappingcnntovmm}
\end{figure}

\subsection{Mapping of convolution layer of LeNet-5 to Vector and Matrix}
\label{lenetmapping}

\par LeNet-5 is a pioneering CNN architecture introduced by Yann LeCun et al. in 1998 for handwritten digit recognition. In this work, we choose LeNet-5 and map the first convolutional layer of LeNet-5 to memory array and evaluate the efficiency of performing VMM by DA. Using LeNet-5 as an example, we demonstrate how the inference of any Neural Network can be executed efficiently as a series of VMM operations in the memory array using our approach. Fig.\ref{mappingcnntovmm} illustrates the mapping of Convolution Layer 1(CONV1) to vector and  matrix. For CONV1, we denote the Input feature as $If_{1}$ and the six output features as $Of_{1,1} \cdot \cdot Of_{1,6}$. CONV1 is fully connected and takes the input image and convolves it with six 5$\times$5 filters to get six 28$\times$28 output feature maps (stride=1, no zero padding). During CONV1, each filter $W_{1},W_{2} \cdot W_{6}$ extracts different low-level spatial features of the input image. During each stride of the convolution, a 5$\times$5 portion of the $If_{1}$ is multiplied with 5$\times$5 filter and summed (inner product or dot product) to calculate one pixel ($Y_{1} \cdot Y_{6}$) of output feature map ($Of_{1,i}$). Since the same $If_{1}$ is multiplied with six different filters, this can be accomplished simultaneously if the filter values (weights) form the columns of the weight matrix. The 5$\times$5 portion of $If_{1}$ during a stride is unrolled into vector [$X_{1} \cdot \cdot X_{25}$]. The 5$\times$5 filter $W_{1}$ is unrolled to column1, filter $W_{2}$ to column2, as depicted in Fig.\ref{mappingcnntovmm}. In this manner, the six 5$\times$5 filters form a 25$\times$6 weight matrix. Each stride of the convolution is therefore a multiplication of a 1$\times$25 vector with a 25$\times$6 matrix resulting in a 1$\times$6 vector. Since a 32$\times$32 image convolved with a 5$\times$5 filter (no padding) produces a 28$\times$28 output requiring 784 convolution operations, CONV1 will require 784 VMMs in memory. Each VMM computes one pixel of six features output feature maps ($Y_{1} \cdot Y_{6}$ being the result pixel of $Of_{1,1} \cdot Of_{1,6}$) and six 28$\times$28 output features maps will be computed after 784 VMM operations.

\subsection{Mapping of Vector and Matrix to memory array}
\begin{figure}[h]
\centering
%\includegraphics[scale=0.55]{./figures/da8by8.pdf}
% converted image.pdf to Press-Ready PDF (PDF/X-4), to fix displaying issues with various pdf viewers
\includegraphics[scale=0.55]{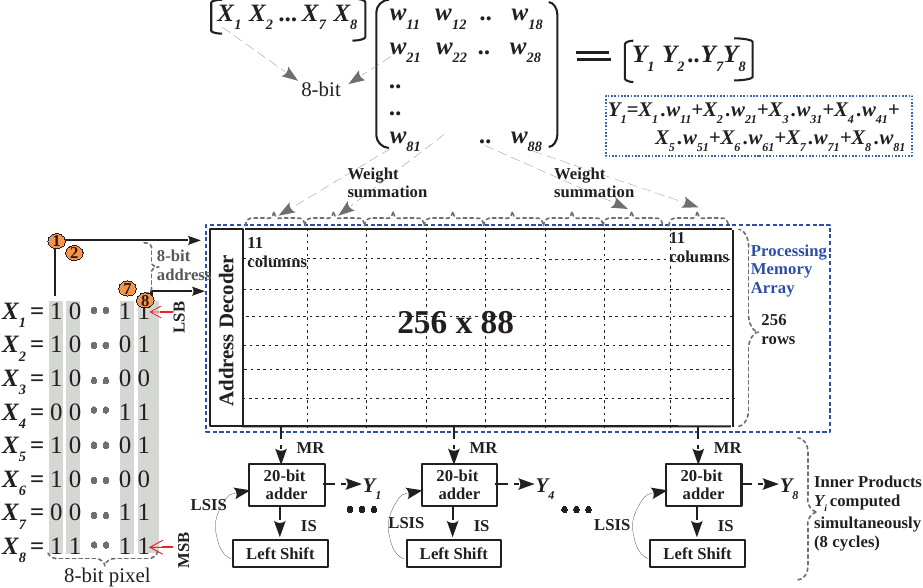}
\caption{\small{A VMM with a 8x8 matrix can be implemented in memory in 8 cycles}}
\label{da8by8}       % Give a unique label
\end{figure}

We trained LeNet-5 on the MNIST dataset using PyTorch. In our implementation, we apply post-training \textbf{symmetric uniform quantization} to convert the trained floating-point weights to 8-bit signed integers (INT8). Research has shown that the inference accuracy of CNN does not degrade much by representing the floating point weights as signed integers ([-128 to 127]) \cite{deeplearninghw}. The input vector, being a grayscale image is a 8-bit unsigned integer([0-255]). For ease of explanation, in this section, we first elaborate how a 1$\times$8 vector is multiplied with 8$\times$8 matrix. For multiplication of vector X with a 8$\times$8 matrix in memory, we will need a 256$\times$88 processing array, as depicted in Fig.\ref{da8by8}. First, the sum of the weights are calculated and written into the columns of the processing array. Since each weight is 8-bit, the maximum possible sum of eight such weights requires 8+$log_{2}(8)$= 11 bits. All possible sum of weights are written to corresponding location of the 256$\times$88 memory $e.g.$ in the first 11 columns of Fig.\ref{da8by8}, at location `10101100', the value of $w_{8,1}+w_{6,1}+w_{4,1}+w_{3,1}$ is written  (corresponding to column 1 of the weight matrix). Similarly, at the same row, the value of $w_{8,2}+w_{6,2}+w_{4,2}+w_{3,2}$ is written in the next 11 columns (corresponding to column 2 of the weight matrix). For 8 columns of a matrix, we need 88 columns of memory to store the sum of the weights. It must be noted that this writing of the sum of weights is a once-in-a-lifetime process and need not be repeated for every inference. Once trained, the CNN weights remain fixed during inference and change only if CNN is re-trained. Hence, although it involves effort (time and energy to write into 256$\times$88), it need not be repeated. In Section \ref{circuitimplementationsection} we elaborate how this effort can be amortized using Non-Volatile Memory.

% fig 4

\par After the weights are written, as depicted in Fig.\ref{da8by8}, the input vector is sliced and in each cycle, 1-bit of [$X_{1},X_{2}, \cdot \cdot ,X_{8} $] is applied to the array's address decoder (Fig.\ref{da8by8}). In each cycle, $MR$ is read out and added with Left-shifted Intermediate Sum ($LSIS$). The MSB of [$X_{1},X_{2} \cdot \cdot ,X_{8} $] forms the address for cycle 1, the next significant bit forms the address for cycle 2 and so on. After 8 cycles, [$Y_{1} \cdot \cdot Y_{8}$] will be available at the output of 20-bit adder. Note that if we had more columns in the weight matrix (say 20 instead of 8), we will still require only 8 cycles to compute [$Y_{1} \cdot \cdot Y_{20}$]. This is because the latency is decided by the bit-width of $X$ and not by the number of columns in the weight matrix. In this manner, VMM is achieved without a hard-wired multiplier usign DA.

% fig 5
\begin{figure}[h]
\centering
%\includegraphics[scale=0.475]{./figures/da16by16.pdf}
% converted image.pdf to Press-Ready PDF (PDF/X-4), to fix displaying issues with various pdf viewers
\includegraphics[scale=0.475]{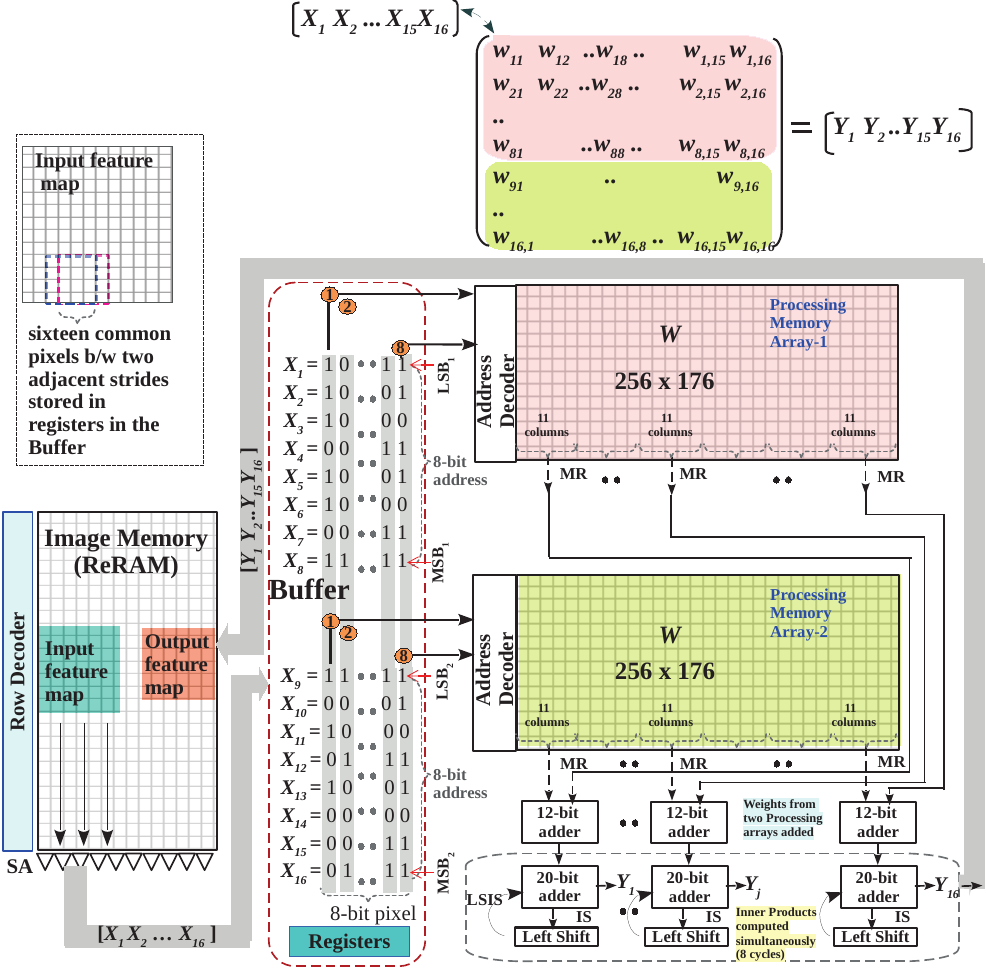}
\caption{\small{Scaling of our approach as the weight matrix scales from 8$\times$8 to 16$\times$16. The 16$\times$16 weight matrix is sliced into two 8$\times$16 matrix and their sum written to two processing arrays. }}
\label{da16by16}       % Give a unique label
\end{figure}
\par What happens if the size of the matrix scales up? For a 16$\times$16 weight matrix, we cannot have a array of $2^{16}$=65536 rows\footnote[2]{Due to the large parasitics of row and column wire,the latency to write and read from arrays increases with its size. Moreover, power consumption increases since it is inefficient to switch on a large array to read and write a small portion of it.}. As depicted in Fig.\ref{da16by16}, we split the processing array into two arrays of 256 rows. The sum of weights can be written in 176 columns (16$\times$11). During VMM, the 16-bit address is split into two 8-bit address and fed to two processing arrays. The weights are read-out and added in a 12-bit adder and then fed to the Add-shift circuit as before. Indeed, the scaling is very good with only one extra adder stage as the size of the matrix scales from 8$\times$8 to 16$\times$16 (Fig.\ref{da16by16}). If the weight matrix is 32$\times$32, we should have 4 such processing arrays and their weights(MR) must be summed together and fed to the add-shift block. Fig.\ref{da16by16} also depicts the overall architecture of performing VMM in memory
using DA. `Image memory' is used to store the input and output feature map. To perform VMM, the portion of the input feature map needed for a single VMM([$X_{1} X_{2} \cdot \cdot X_{16}$]) is read out of the image memory and fed to the Processing Memory Arrays(PMA-1 and PMA-2). The buffer serves to store $X$ temporarily and feed one bit of [$X_{1} X_{2} \cdot \cdot X_{16}$] in each cycle. Additionally, the registers in the buffer are used to temporarily store the 16 pixels common between adjacent strides of the convolution. After 8 cycles, the result of VMM [$Y_{1} Y_{2} \cdot \cdot Y_{16}$] is written to the image memory again. Thus, the inputs $X$ and outputs $Y$ of VMM are only moved between adjacent arrays, conserving energy. Furthermore, all the adders (12-bit adder used to add the sum of weights from PMAs and 20-bit adder) are high performance Ladner-Fischer parallel-prefix adders which are fast and energy-efficient \cite{tvlsi}.

\section{Circuit Implementation of VMM}
\label{circuitimplementationsection}
% fig 6
\begin{figure}[h]
\centering
%\includegraphics[scale=0.5]{./figures/preconvolutionvmm.pdf}
% converted image.pdf to Press-Ready PDF (PDF/X-4), to fix displaying issues with various pdf viewers
\includegraphics[scale=0.5]{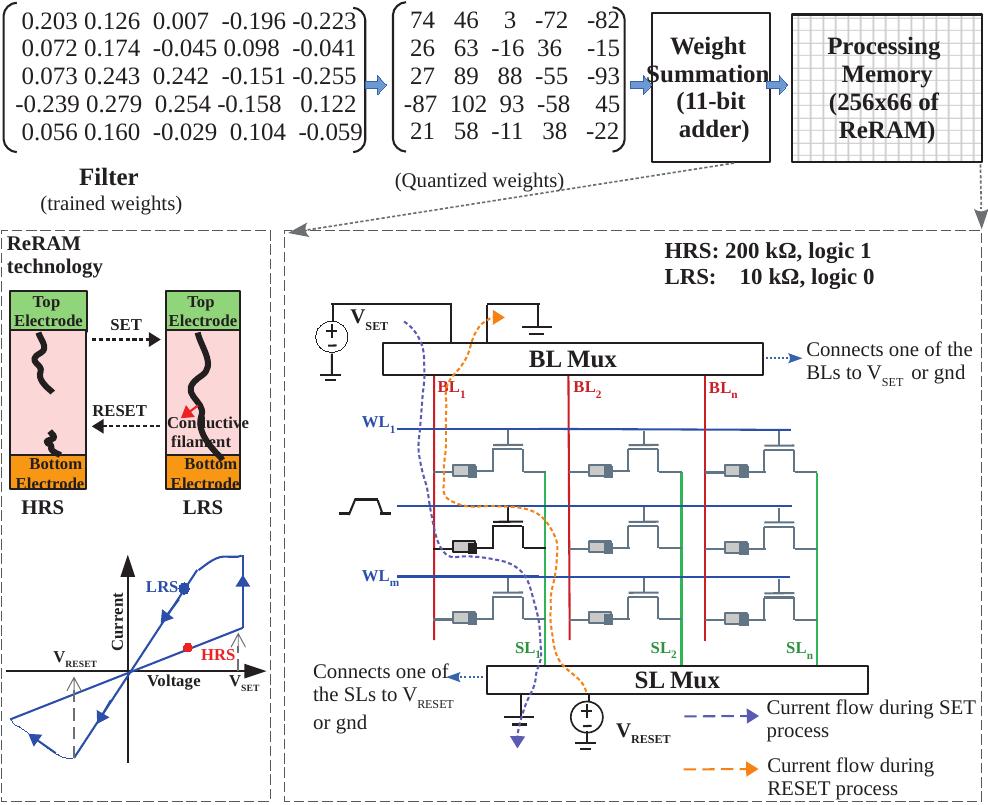}
\caption{\small{Before performing VMM, the co-efficients of the filter are summed and written to the processing memory made of ReRAM cells. }}
\label{preconvolutionvmm}       % Give a unique label
\end{figure}

In this section, we implement the multiplication of [$X_{1} X_{2} \cdot \cdot X_{25}$] with 25$\times$6 weight matrix of CONV1 of LeNet-5. Our purpose is to functionally verify our DA approach by transistor-level simulation and also calculate the energy and latency for a single VMM. Our choice of memory technology was dictated by the need to amortize the cost of writing (energy and time) the sum of the weights into the memory. A volatile memory could be used for processing memory but will incur cost every time we want to do inference (convolution with the same filters). So, we chose Non-Volatile memory technology. Among NVM, we chose Resistive RAM due to it's simple programming circuits (does not need high voltage WRITE circuits like flash memory) and maturity. Resistive RAMs (ReRAMs) are two terminal devices (usually a Metal-Insulator-Metal structure) capable of storing data as resistance. When subject to voltage stress, it’s resistance can be switched reversibly between a Low Resistance State (LRS) and a High Resistance State (HRS) (Fig.\ref{preconvolutionvmm}). The change of resistance is due to the formation or rupture of a conductive filament in the insulator. Typically, ReRAM is fabricated in a 1T-1R structure $i.e.$ each memory cell has an access transistor to avoid sneak-path currents during reading and writing. Before, we elaborate how VMM of CONV1 layer is performed in memory, we briefly describe the pre-VMM procedure we followed to write the weights into the processing array.

\subsection{Pre-VMM procedure}
\label{preconvolutionsection}
As depicted in Fig.\ref{preconvolutionvmm}, the weights from training the LeNet-5 are floating point. They are quantized to 8-bit signed integers and fed to weight summation adder (11-bit adder) to calculate the sum of the weights. The sum of the weights are then written to the 1T-1R Resistive memory. When a positive voltage, $V_{SET}$, is applied, the device switches from HRS to LRS. When a voltage of opposite polarity, ($V_{RESET}$), is applied, the devices switches back to HRS. In 1T-1R array, SET process is accomplished by applying $V_{SET}$ to the top electrode($BL$) and grounding the source terminal and for the RESET process, the polarity is reversed. Denoting HRS as logic `1' and LRS as logic `0', a 11-bit value (10101100111) can be stored as (HRS LRS HRS LRS HRS HRS LRS LRS HRS HRS HRS) in the memory, where HRS is stored by a RESET operation and LRS is stored by a SET operation at corresponding locations in the memory. In this manner, the sum of weights are stored in a non-volatile memory. Once written, the sum of the weights remains unaltered in the processing memory (typical retention of Resistive RAM is 10 years) and any number of VMM can be performed on the same weight matrix with different input vector (input image). Since the weights are fixed after training, our in-memory architecture is ideally suited for inference of CNNs.

\begin{figure}[h]
\centering
%\includegraphics[scale=0.5]{./figures/da25by6.pdf}
% converted image.pdf to Press-Ready PDF (PDF/X-4), to fix displaying issues with various pdf viewers
\includegraphics[scale=0.5]{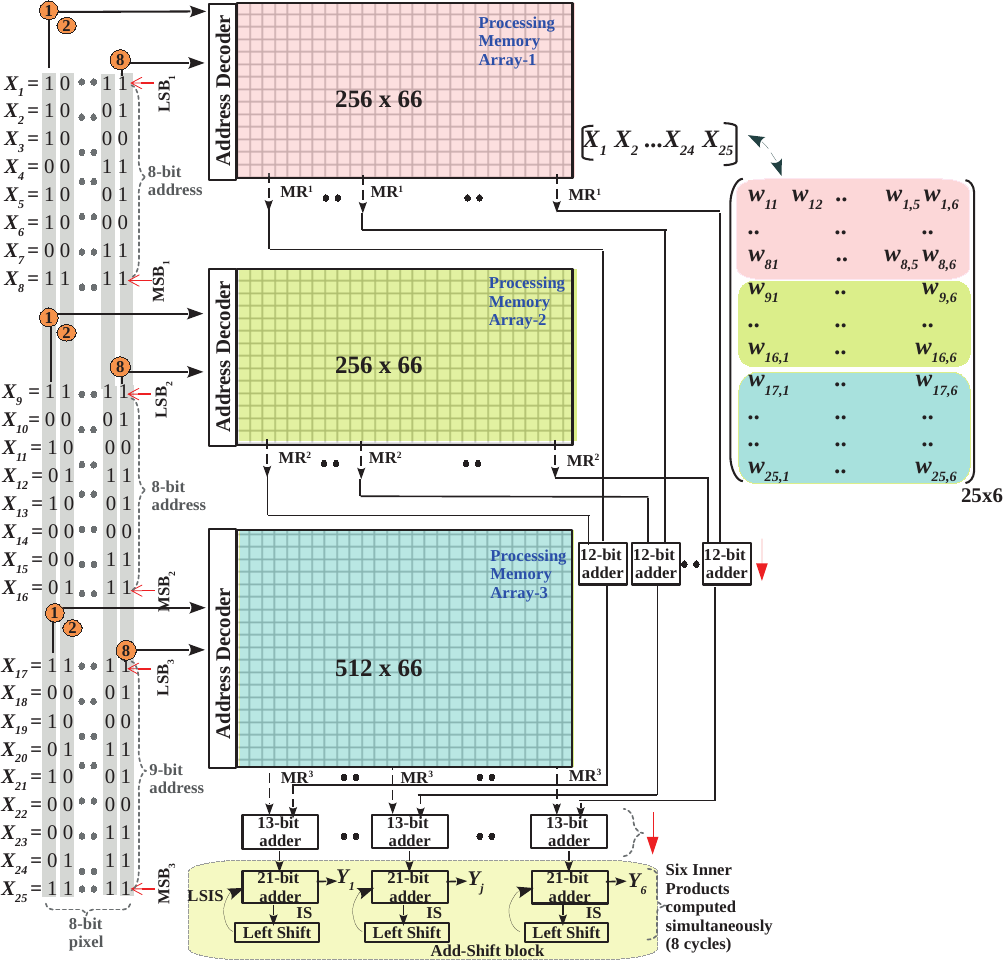}
\caption{\small{CONV1 of LeNet-5-- multiplication of 25-element input vector with a 25$\times$6 matrix is performed in three processing arrays. }}
\label{da25by6}       % Give a unique label
\end{figure}

\subsection{Reading Methodology from ReRAM Array}
\label{readsection}
\begin{figure}[h]
\centering
%\includegraphics[scale=0.65]{./figures/savmm.pdf}
% converted image.pdf to Press-Ready PDF (PDF/X-4), to fix displaying issues with various pdf viewers
\includegraphics[scale=0.7]{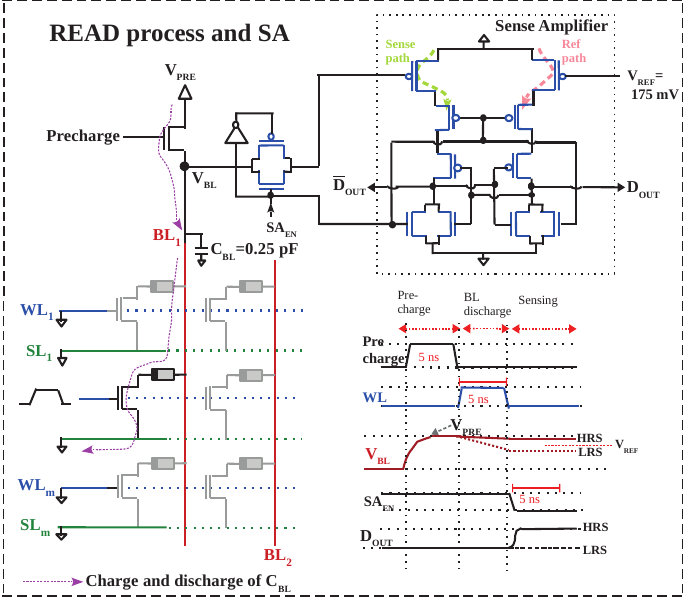}
\caption{\small{READ process and Sense Amplifier. Sensing is performed in 15 ns by a sequence of `Pre-charge',$WL$ and $SA_{EN}$ signals, each activated for 5 ns. }}
\label{savmm}       % Give a unique label
\end{figure}

Since the time and energy to read from the Image/Processing memory array significantly affects the efficiency of our VMM (single VMM has 8 READ from memory array followed by add and shift), we optimized our SA for efficient reading. Our sensing technique is based on pre-charging the $BL$ to a specific voltage and then discharging it through the memory cell to be sensed\cite{rerambook}. After discharging, the bit line voltage $V_{BL}$ is compared with a reference voltage $V_{REF}$ to sense the state of the cell (Fig.\ref{savmm}). We assumed the bit line capacitance of our memory array to be 250 fF which is the reported $BL$ capacitance for a similar memory array fabricated in 130 nm process \cite{feramjournal}. After precharge and discharge (Fig.\ref{savmm}), $V_{BL}$ is compared with a reference voltage, $V_{REF}$ using a Sense Amplifier(SA). A Transmission Gate (TG) is used to feed this $V_{BL}$ to the PMOS of the comparator. The purpose of the TG is to decouple the $BL$ electrically after $WL$ goes LOW because $BL$ must be pre-charged for the next READ operation (this is necessary for pipelining, as described in next Section). The energy consumed during READ is calculated to be 35 fJ  (more details of the sensing process can be found in \cite{rerambook}).

% fig 8

\subsection{In-Memory Vector Matrix Multiplication Circuit}
 As depicted in Fig.\ref{da25by6}, VMM of CONV1 layer of LeNet-5 can be implemented in 3 processing arrays. The 25$\times$6 matrix is sliced into three sub-matrices and mapped to three processing arrays. Processing Memory Array-1 (PMA-1) contains all possible sum of weights corresponding to [$w_{1,i} \cdot \cdot w_{8,i}$], where i $\in$ 1-6 filters. Similarly, PMA-2 and PMA-3 contains all possible sum of weights corresponding to [$w_{9,i} \cdot \cdot w_{16,i}$] and [$w_{17,i} \cdot \cdot w_{25,i}$], respectively. The weights are written to three PMAs using the procedure described in section \ref{preconvolutionsection} and this is a once-in-lifetime procedure. The vector $X$ is also sliced into three sections -- [$X_{1} \cdot \cdot X_{8}$], [$X_{9} \cdot \cdot X_{16}$] and  [$X_{17} \cdot \cdot X_{25}$] and applied as address for PMA-1,PMA-2 and PMA-3, respectively. Note that vector [$X_{17} \cdot \cdot X_{25}$] is a 9-bit vector and hence PMA-3 is a 512$\times$66 array.
\par The entire VMM is performed in 8 memory cycles. In each cycle, 1-bit of [$X_{1} X_{2} \cdot \cdot X_{25}$] is applied to three PMAs simultaneously. $MR^{1}$,$MR^{2}$ and $MR^{3}$ denote the \textbf{M}emory \textbf{R}ead-out from PMA-1,PMA-2 and PMA-3. $MR^{1}$ and $MR^{2}$ are first added using a 12-bit  adder (adding two 11-bit signed integers requires at least 12-bit adder to prevent overflow). The 12-bit result ($MR^{1}+MR^{2}$) is added with $MR^{3}$ in a 13-bit adder to get the final sum of the weights (this would be the actual sum of weights we would have read out if we had an array of size $2^{25}$$\times$66). This final sum of weights is fed to the Add-and-Shift circuit as before. After 8 cycles, the result [$Y_{1} \cdot \cdot Y_{6}$] will be available at the output of 21-bit adders.
[$Y_{1} \cdot \cdot Y_{6}$] is one pixel of the 6 output feature maps and the process is repeated 784 times to compute the complete CONV1 layer.
% fig 9
\begin{figure*}[h]
\centering
% converted image.pdf to Press-Ready PDF (PDF/X-4), to fix displaying issues with various pdf viewers
% trim=left bottom right top
\includegraphics[scale=0.35,trim=1cm 13cm 9cm 1cm,clip]{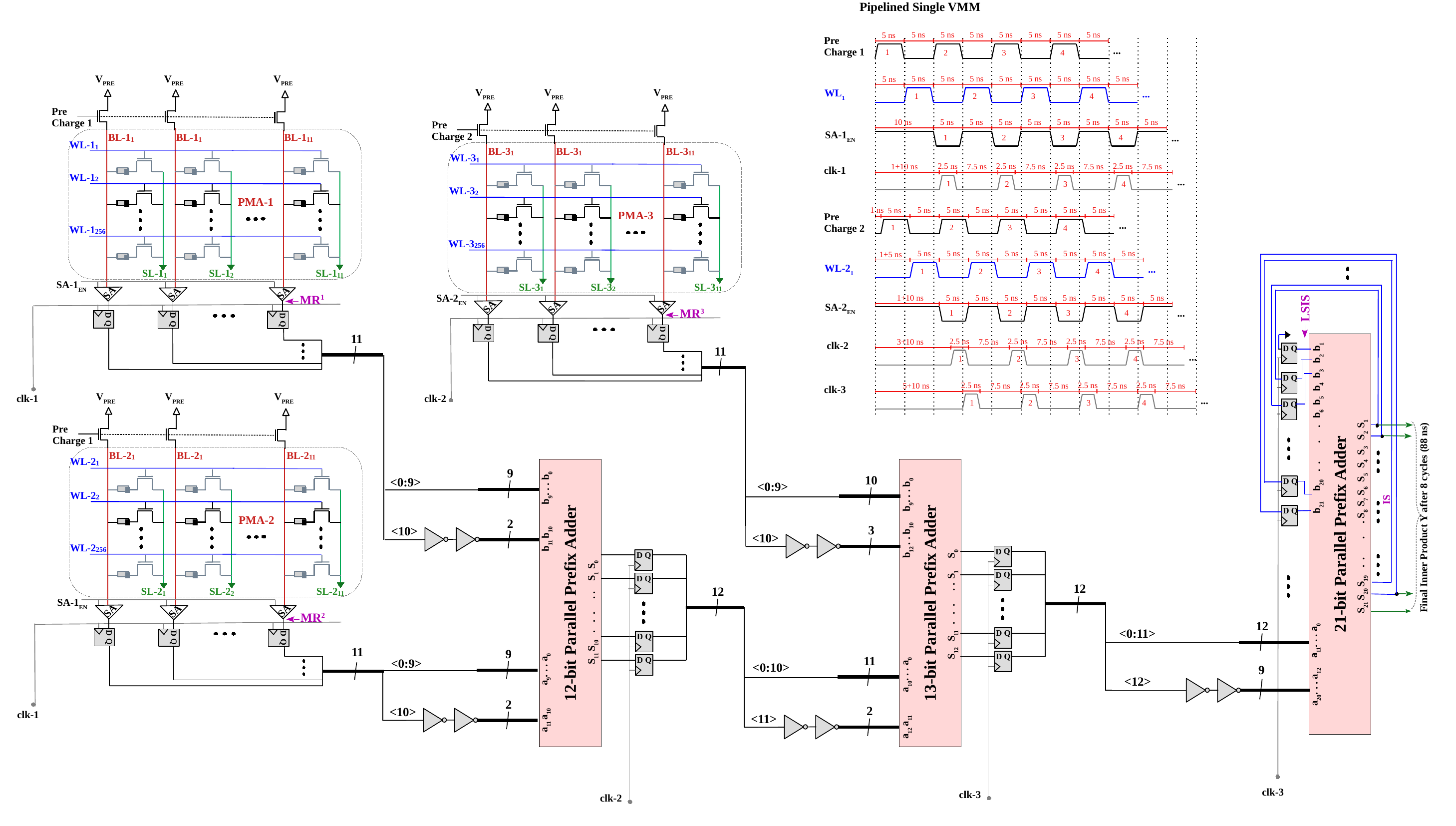}

\caption{\small{Detailed circuit to implement VMM: Due to space constraints, we have shown the circuit performing the VMM for one column of the matrix of Fig.\ref{da25by6}. ($a_{20} \cdot \cdot a_{0}$) forms the first input to the 21-bit adder and is fed with $MR$. ($b_{20} \cdot \cdot b_{0}$) is the second input of the adder and is fed with LSIS. For inference using quantized weights (Fig.\ref{preconvolutionvmm}), weights are 8-bit signed integers and the sum of weights (11-bit) is read out of processing memory per column of weight matrix. The remaining 10-bits ($a_{20} \cdot \cdot a_{10}$) are initialized to either `1' if left-most bit of $MR$ is `1' (negative weight) or `0' if left-most bit of $MR$ is `0' (positive weight).}}
\label{vmmcircuit}
\end{figure*}

Fig.\ref{vmmcircuit} depicts the detailed version of Fig.\ref{da25by6} we simulated. The outputs $MR^{1}$ and $MR^{2}$ from PMA-1 and PMA-2 are first added in the 12-bit adder at the rising edge of clk-1 (t=$10+1$ns). At the rising edge of clk-2, ($MR^{1}+MR^{2}$) is added with $MR^{3}$. Reading out PMA-3 and clk-2 is delayed by 2 ns (t=$11+2$ns) to give time for addition of $MR^{1}$ and $MR^{2}$. At clk-3 after an additional delay of 2 ns (t=$13+2$ns), the 21-bit adder accumulates the result from the three PMAs and the left-shifted result from the previous cycle. The VMM operation is accomplished through this pipelined architecture where each of the 8 cycles processes one bit of the input vector and its result accumulated in the 21-bit adder with left-shift operations (to account for multiplying of IS by 2). After 8 cycles, the result $[Y_1...Y_6]$ is available at the outputs of the six 21-bit adders.
\\

\subsection{Energy and Latency for VMM}

The Energy of the single VMM of Fig.\ref{da25by6} was determined by integrating the current drawn from $V_{DD}$ over 8 cycles and found to be 110.2 pJ. The total latency for a single VMM is 88 ns. The VMM starts with a first cycle of 15 ns consisting of Precharging (5 ns), Discharging (5 ns) and Sensing (5 ns). Sensing (of a particular set of data) and Precharging (in preparation for the next set of data to be read out) can be done in parallel due to the Sense Amplifiers electrical decoupling of the bitline, which saves 5 ns for each consecutive cycle. After 7 consecutive cycles of 10 ns each and a last 21-bit addtion ($<3$ns), a single VMM is completed in 88 ns (15 ns + 70 ns+ 3 ns).\\
\par Note that 110.2 pJ is the energy to compute VMM and does not include the energy consumed during pre-VMM procedure. We need to write the sum of the weights to three processing arrays (two 256$\times$66 and one 512$\times$66). First, we need to calculate the sum of the weights and then write them in to the PMAs. Since this is not time critical (once in a lifetime process), we assume a single accumulator (11-bit adder) which can add the weights serially. For the three PMAs of Fig.\ref{da25by6}, this will require 24576 additions, where each addition consumes an energy of 52 fJ.$E_{weightsummation}$=24576 $\times$ 52 fJ=1.27 $nJ$. To that, we add the energy to write in to the ReRAM array, typically 1 pJ/bit\footnote{Note that we are writing only bits into ReRAM which requires programming the device to HRS/LRS and does not require complicated programming algorithms. In contrast, writing multi-bit requires algorithms like ISPVA increasing WRITE energy per bit.}. $E_{write}$= 67584$\times$1 pJ= 67.58 nJ and the total energy (weight summation+writing into PMA) is 68.8 nJ. It must be noted that this energy is once-in-a-lifetime cost. This energy is amortized over the number of times the memory array is used for VMM (Inference). Even if we consider a moderate 10000 inferences during the lifetime of the ReRAM chip, this adds an energy of only 6.88 pJ per inference. 

\section{Comparison with Bit-slicing approach}
\label{comparisonsection}
\begin{figure*}[h]
\centering
\includegraphics[scale=0.65]{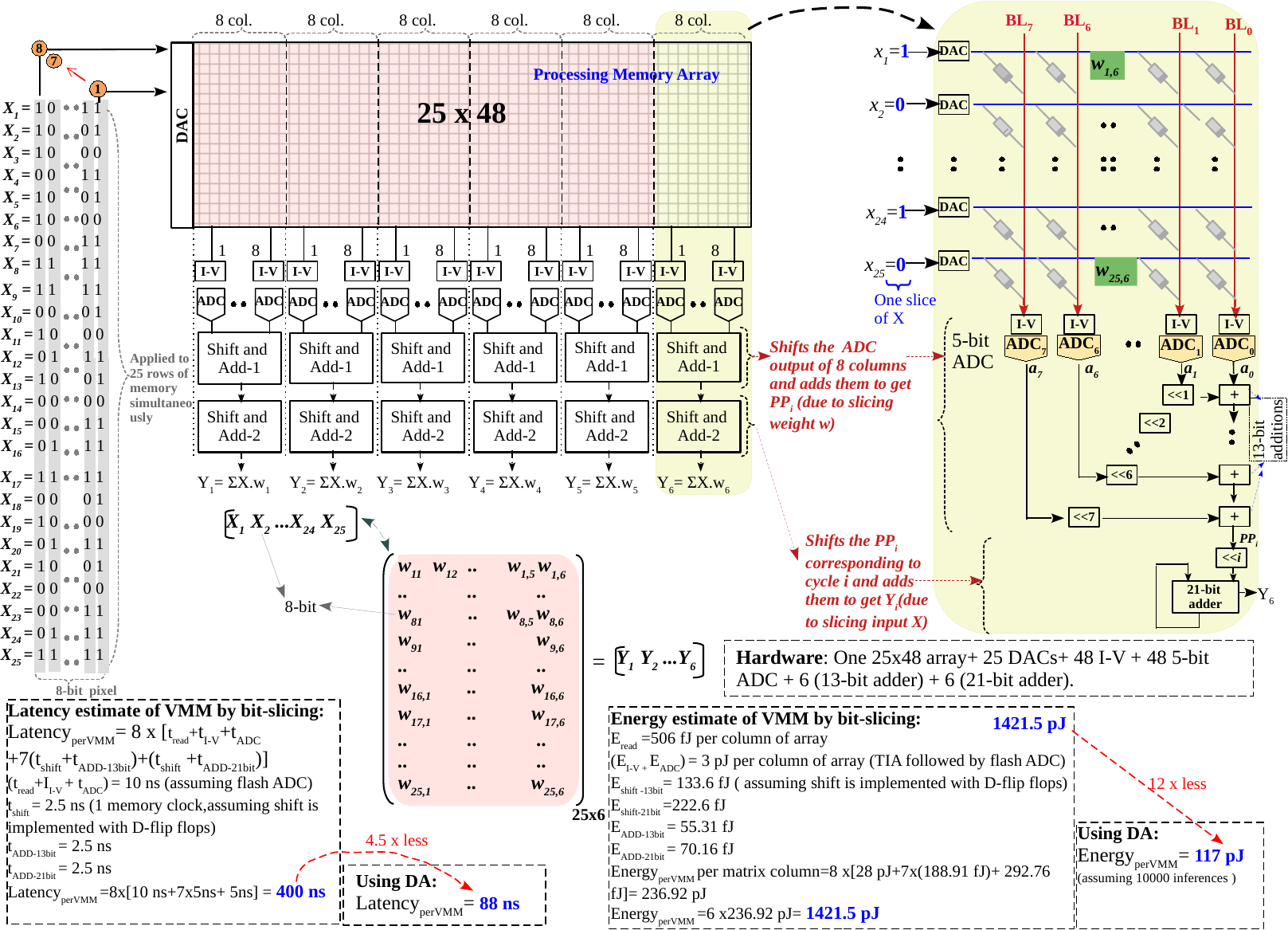}
\caption{Illustration of Vector (size 1$\times$25) matrix (size 25$\times$6) multiplication operation by bit slicing approach}
\label{vmmcomparison_bitslicing}
\end{figure*}

The most common way of performing VMM in memory in the present literature is by bit-slicing \cite{issac,enhancedbs}. To overcome the high resolution ADC requirement\footnote{an ADC with higher resolution requires more area and consumes more power}, other in-memory VMM works store the weights in binary form and multiply the output of each column by appropriate power of 2 outside the array. In essence, they are `slicing' the `weight' over different columns of the array. The inputs are also sliced and fed to the array using a simpler DAC over 8 cycles in a bit-serial manner. As depicted in Fig. \ref{vmmcomparison_bitslicing}, the same VMM can be implemented in one 25$\times$48 array with dedicated ADCs and Shift and Add circuits. The 8-bit weights are stored in 8 columns, requiring 48 columns for the 6 columns of the weight matrix. Eight columns of the array are zoomed out on the right to illustrate the VMM process better. The 25-element vector is sliced and in each cycle, one bit of $X$ is applied to the DACs, starting from the LSB. The DAC applies the bit as corresponding voltage to the $WL$, a `1' as 0.2 V and `0' as 0 V. This is followed by the READ operation, where the corresponding current in each memory location are added along the columns resulting in $BL$ current. The $BL$ current is converted to a voltage using an I-V converter followed by a 5-bit ADC\footnote{because the minimum and maximum decimal value of sum of 25 rows can be 0 and 25 respectively}. The digital output of the eight ADCs are processed using the two Shift-and-Add circuits. The purpose of the first Shift-and-Add is to undo the slicing of the weight `w' and purpose of the second Shift-and-Add is to undo the slicing of the input `X'. For fair comparison, we estimated the latency and energy for this bit-slicing approach with the same ReRAM technology and the same shifters and adders we used for DA (Fig.\ref{vmmcircuit}). Since shifting is implemented using D-flip flops, we assume 2.5 ns for the shift operation. The ADD operations are also clocked and require 2.5 ns like the ADD operations in Fig. \ref{vmmcircuit}. The total latency for a VMM is 400 ns using bit-slicing while our DA approach incurs only 88 ns.

\par The energy $E_{read}$ is due to the current flow in the $BL$s and is estimated to be 506 fJ\footnote{$E_{read}$ is estimated by assuming the same HRS and LRS value. On an average 12 out of 25 rows can be in LRS and the remaining in HRS. $I_{READ}$ is therefore 253 $\mu$A. Assuming $V_{READ}$ of 0.2 V and $t_{READ}$ of 10 ns, $E_{READ}$=0.2 V$\cdot \int_{0}^{t_{READ}} I_{READ}.dt $= 506 fJ.}. $E_{I-V}$ together with $E_{ADC}$ is estimated to $\approx$ 3 pJ (estimated for 130 nm process). The total energy for a VMM is estimated to be 1421.5 pJ while the proposed DA consumes 117 pJ (Fig.\ref{vmmcomparison_bitslicing}). Table \ref{summary} compares the quantitative latency, energy and area requirement of bit-slicing approach with the proposed DA approach. In summary, the latency is 4.5 $\times$ less and the energy is 12 $\times$ less  compared to the bit-slicing approach. It must be noted that a SA used in our DA is a simple comparator which occupies much less area compared to a 5-bit ADC\footnote{As a quantitative comparison, a 5-bit flash ADC requires 31 comparators (each having at least 9 transistors) followed by a thermometer-to-binary circuit (400 transistors) resulting in 679 transistors and 32 resistors per ADC.} for each column. The DA approach understandably requires large memory to store the sum of weights. However the peripheral hardware required is still lesser than that used in bit slicing due to the ADC/DACs.

\begin{table}[h]
   % \centering
     \caption{Comparison of Energy, latency, hardware required for 1$\times$25 vector with 25$\times$6 matrix multiplication}
    \begin{tabular}{p{1cm}|p{2cm}|p{2.3cm}|p{2cm}}
    \hline
         & Bit-slicing & DA & DA is  \\
         \hline
        Latency & 400 ns & 88 ns & 4.5$\times$ less\\
        \hline
        Energy & 1421.5 pJ & 110.2 pJ+6.88 pJ$^{*}$=117 pJ &  12$\times$ less\\
        \hline
        Hardware & 25$\times$48 array + 25 DACs + 48 I-V + 48 5-bit ADCs + 6 13-bit adder + 6 21-bit adder &  Two 256$\times$66 and one 512$\times$66 array + 0 DACs + 198 SA + 6 12-bit adder + 6 13-bit adder + 6 21-bit adder & \\
        \hline
        Area (transistor count) & 1200 memory cells + 47286 transistors$^{**}$ + 1584 resistors$^{***}$  & 67584 memory cells + 20622 transistors & 56$\times$ more memory cells, 2.3$\times$ less transistors and no passive resistors\\
        \hline
    \end{tabular}
    \label{summary}\\
    $^{*}$ $E_{preVMM}$ per inference assuming 10000 inferences during lifetime\\ 
   $^{**}$Transistor estimation assumes transmission-gate based 2:1 Mux for DAC, Flash architecture for ADC, op-amp based transimpedance amplifier for I-V.\\
   $^{***}$ resistors of flash ADC and I-V converter.
\end{table}

\section{Conclusion}
\label{conclusionsection}
In this work, we have presented a novel method to perform VMM in memory which overcomes the two hurdles of existing techniques to perform VMM in memory-- the requirement of power-hugry ADC and the requirement to program the memory cell to multiple states. It must be noted that even bit-slicing approach requires an ADC (albeit smaller compared to pure analog VMM) while our approach requires an ordinary SA (to read bits) which is an integral part of any memory array. Hence, we re-purpose a normal memory to perform VMM with minimal alterations to the peripheral circuitry $i.e.$ extra adder circuits to aid in computation.  The significant contribution is that using distributed arithmetic, the need for power-hungry ADC/DACs is avoided and our approach uses a simple SA which is integral part of any memory and not power hungry. Therefore, our methodology can be adopted by existing memory technologies and adders are all you need to compute VMM in memory. Our approach is endurance-friendly since we use the ReRAM processing array as a `Read Only Memory'. Finally, our method is `almost digital' requiring only CMOS-based SA and adders and consequently easily scalable down to lower process nodes unlike other approaches which rely heavily on ADC/DAC (and therefore not scalable) for VMM in memory.

\section*{Acknowledgment}
\small{This research was performed in the Bavarian Chip Design Center (BCDC) funded by Bayerisches Staatsministerium für Wirtschaft, Landesentwicklung und Energie }

\bibliographystyle{IEEEtran}
\bibliography{references}
\end{document}